# Classification of genetic variants using machine learning

Abhinav Jain, Greg Slabaugh, Deepti Gurdasani
School of Electronic Engg. & Computer Science
Queen Mary University of London, UK

*Abstract*—Recent advances in genomic sequencing technology have resulted in an abundance of genome sequence data. Despite the progress in interpreting those data, there remains a broad scope for their translation into clinical and societal benefits. Loss-of-function variations in the human genome can be causal in disease development. Precise identification of such variations and pathogenicity prediction may lead to better drug targeting, among other benefits. Machine learning comes across as a promising method for its proven predictive ability. We have curated a novel dataset for the classification of LOF variants using high-quality databases of genetic variation. We trained and validated seven different classification algorithms using the new dataset to classify the variants as Benign, Pathogenic and Likely pathogenic. We recorded the best overall performance using the XG-Boost algorithm with an F1-score of 0.88 on the test set. We observed fair performance on Pathogenic samples with high recall and moderate precision and subpar performance on Likely pathogenic class, albeit with moderate precision. Overall, the encouraging results make our final model a promising candidate for further real-world tests.

## I. INTRODUCTION

The advent of next-generation sequencing (NGS) and a significant drop in the genomic sequencing cost, from a billion dollars per genome in the early 2000s to less than a thousand dollars in 2020 (NGHRI, 2020), have propelled the phenomenal growth of genome sequence data. However, this abundance of data has not proportionately translated into clinical research outcomes such as drug discovery and prioritization due to interpretability and methodological challenges.

Variations in the human genome cause all the diversity in humans, like race and appearance. Although most of the variations are believed to be neutral, some may cause or influence the risk of common and rare diseases, for instance, the association of colorectal cancer risk with the variations in the three genes as demonstrated by Dunlop et al. (2012).

Loss-of-function (LOF) variants in the human genome are low-frequency variations that may significantly alter or inactivate the function of protein-coding genes and thus can be potential drug targets. However, LOF variants are also found in many healthy individuals and are susceptible to false positives (Karczewski, et al., 2020). Thus, high quality and rigorous mechanisms are required for calling and determining the clinical significance of LOF variants.

Mapping between genetic variation and clinical significance is still far from fully understood as the genome sequence of only a tiny proportion of the human population is hitherto available. It makes the mapping extremely hard and realistically unviable to code and provides an opportunity for machine learning methods to draw insights from the curated data and empirical results. Several researchers have used machine learning methods on genetic variations such as Single Nucleotide Polymorphism (SNP) and missense variants; however, very little work has been done using LOF variants.

We have created a novel dataset by extracting and combining data from high-quality databases, viz. gnomAD (LOF variants), Clinvar (clinical significance), Ensembl (variant effects) and East London Genes and Health (ELGH, healthy individuals). We used this dataset to train and evaluate various machine learning classification algorithms to classify LOF variants as Pathogenic, Likely Pathogenic and Benign. We achieved the best F1-score of 0.88 with the tuned XGBoost algorithm. The predictions by the final model look promising to use alongside existing methods in clinical tasks such as drug target prioritization to test the model in a real-world setting. The model's predictions may function as a second opinion to the existing methods of prioritizing the drug targets.

All the source code, models and most data excluding the restricted components are available at our project's public [GitHub repository](#).

## II. RELATED WORK

The availability of whole-genome sequencing over the last two decades has accelerated the study of genetic variation across species. Internationally coordinated efforts and publicly available databases such as 1000-genome (The 1000 Genomes Project Consortium, 2015), Clinvar (Landrum, et al., 2016), gnomAD (Karczewski, et al., 2020), dbSNP (Sherry, et al., 2001), ENCODE (ENCODE Project Consortium, 2012), GWAS (Welter, et al., 2014), HGMD (Stenson, et al., 2017) and SWISS-PROT (Bairoch & Apweiler, 2000) have enabled the progress of research into the clinical significance of different kinds of genetic variants.

Despite the enormous progress, the field of genetic variation still has more questions than answers. The complex data and poor interpretability have inspired several machine-learning driven projects to enhance understanding of the genetic variation effects.

Bendl et al. (2014) created PredictSNP, a consensus classifier for predicting the effects of Single Nucleotide Variants (SNVs) on protein function, combining six leading classifiers at that time. They used an integrated protein-focused dataset without duplicates from the six projects and trained seven different classification algorithms. They used an ensemble algorithm based on the majority vote weighted by confidence to derive their models' prediction. PredictSNP exceeded the performance of all individual models; however, there was no noteworthy difference between the seven classification algorithms used.

Combined Annotation Dependent Depletion (CADD), a framework developed by Kircher et al. (2014), integrated diverse annotations for SNVs and small indels and provided a measure of deleteriousness of genetic variants applicable to any possible human SNV. Kircher et al. (2014) sourced

Email addresses:
| abhinav.jain@se20.qmul.ac.uk | g.slabaugh@qmul.ac.uk | d.gurdasani@qmul.ac.uk |

annotations from Ensembl's Variant Effect Predictor (VEP), UCSC genome browser tracks, GERP conservation score, Grantham, SIFT and PolyPhen. They ended up with 29.4 million variants and 63 distinct annotations used as predictors for training the models. They further trained ten Support Vector Machines (SVMs) with linear kernel and derived an average of the ten models to calculate the CADD score for all SNVs and indels in scope. Kircher et al. (2014) claimed the CADD score to be a more effective measure of deleteriousness than the existing, inevitably biased, incomparable, and diverse annotations. All the calculated CADD scores are publicly available. Although a good measure of the deleteriousness that can link to the pathogenicity of the variants, the CADD score is currently available only for SNVs and small indels, and the framework requires expansion to calculate CADD scores for all available loss-of-function variants.

Quang et al. (2015) followed up the original CADD framework with DANN, in which they replaced the SVM model with a three-hidden-layered neural network. They used the same dataset as Kircher et al. did in the CADD and reported an improved performance over the original CADD. However, DANN offered nothing more than CADD in terms of variant effects coverage.

PredictSNP2 (Bendl, et al., 2016) resonated with its predecessor PredictSNP (Bendl, et al., 2014) in using the consensus philosophy, albeit with different datasets predominantly using variant-based predictors. Bendl et al. (2016) created three datasets characterized by the associated diseases. They used Clinvar as the source of variants linked with the Mendelian diseases, NGHRI GWAS catalog for complex diseases and COSMIC for cancer-linked variants. They supplemented all the datasets with neutral variants from the VariSNP database. For evaluating consensus, they only polled the databases that had precalculated scores for the variants in scope. It significantly restricted the generalizability of predictSNP2, downgraded the credibility of its predictions for variants without a precalculated score in the intrinsic databases, and rendered predictSNP2 a mere aggregator of the available scores.

Pagel et al.'s (2017) MutPred-LOF was perhaps one of the first projects focused on LOF variants and their classification based on machine learning. Pagel et al. created a dataset of 130529 variants that had ~24% pathogenic variants. They obtained pathogenic variants from Clinvar and Human Gene Mutation Database (HGMD) and neutral variants from gnomAD's ExAC database. Their main idea for classifying LOF variants was based on the consideration of the protein function, and they emphasized the features based on the protein sequence. MutPred-LOF was designed as a 100-bagged feed-forward neural network with two hidden layers and was trained using the resilient-propagation algorithm. The majority class was subsampled for balancing the training data. MutPred-LOF demonstrated adequate performance on the validation set and provided insights into the characteristics of LOF variants, such as the interplay between variant and protein-based features. However, its utility remained restricted due to shortcomings in the available data (Pagel, et al., 2017). Nevertheless, it showed that computational and machine learning methods could be effective in studying the LOF variants.

DeepPVP (Boudellioua, et al., 2019) relied on the idea of combining pathogenicity with phenotype similarity for the prioritization of causative variants. Boudellioua et al. used nearly 50,000 pathogenic and benign variants from Clinvar along with annotations from CADD, DANN and Genome-wide Annotation of Variants (GWAVA) to obtain their dataset. They trained a neural network and exceeded the training performance of DeepPVP's predecessor; however, DeepPVP performed poorly upon testing with a test set from the 1000-Genome database. Ignorance for categorizing variants based on functional consequence and the employed annotations might have contributed to its inadequate generalizability.

Despite many promising machine learning projects dealing with genetic variation data, establishing a concrete framework remains challenging due to genomic data's evolving nature and interpretation. Creating new datasets, upgrading existing ones, creating or updating learning algorithms seem inevitable to keep up with the evolution of genomic interpretation and machine learning techniques.

Our work makes two main contributions, first a novel dataset for LOF variants intended for machine learning classification tasks constructed from high-quality data sources and second, an evaluation of various classification algorithms for performing the classification task on our new dataset.

### III. METHODOLOGY

#### A. Dataset

Genome Aggregation Database (gnomAD) project hosts a rich aggregation of genome sequencing data from various leading international projects (Broad Institute, 2020). Some of the constituents in its various publicly available data repositories include genome and exome variations, mutational constraints, multi-nucleotide variants, structural variants and LOF curation results. It is also the largest high-quality database of LOF variants (Karczewski, et al., 2020). It has LOF variants identified using the combination of VEP annotation and further filtering by LOF Transcript Effect Estimator (LOFTEE).

Using the LOF curation results available on the gnomAD portal (All homozygous, Lysosomal storage disease genes, AP4, FIG4, MCOLN1, Haploinsufficient genes and Metabolic conditions genes), we identified 1894 genes relevant to LOF variants. gnomAD portal provides a function to download all variants of a given gene. The downloaded variant data contains all Variant Calling Format (VCF) fields and, in addition, includes several important features such as Clinvar's clinical significance, VEP annotation, allele frequency distribution among various ethnicities, homozygote count and hemizygote count. However, the function only works for one gene at a time.

We developed a Java program to automate en masse extraction of all the genes of our interest. The program is available on our project's public [GitHub repository](#). Using this program, we downloaded 1894 files and then combined them into a single file using a Linux shell command. We configured the extraction program to select only the LOF variants option before the download. The format of variant data downloaded from gnomAD acts as the backbone of our dataset's schema.

The initial extraction resulted in 678654 variants; however, Clinvar's clinical significance was available only for 11980 variants; hence, we discarded all the variants without any clinical significance. Clinvar is a manually curated, freely available database where clinical practitioners from all over the world submit their interpretations of the clinical significance of the variants (Landrum, et al., 2016). We further filtered the data with clinical significance as Pathogenic, Likely pathogenic and Pathogenic/Likely pathogenic and ended up with 2020 variants.

Loss-of-function observed/expected upper bound fraction (LOEUF), a metric for enabling quantitative assessment of constraints (Karczewski, et al., 2020), available precalculated on gnomAD portal, was downloaded and merged with our data using the transcript id as the shared key, which further filtered the data down to 2012 records as the metric was not available for the eight variants.

The data extracted so far only contained disease-related variants. We further used the East London Genes and Health's (ELGH) database for supplementing our data with the variants found in the healthy population. ELGH's database contains genomic data of more than 100000 Bangladeshi and Pakistani origin people living in East London and Bradford (Genes & Health, 2021). The data is not publicly available and requires download access approval from the concerned authorities. ELGH data contained 29.58 million records/9.86 million variants found in healthy people and people with complex diseases such as diabetes. We sampled 5000 records from the ELGH database, keeping the chromosome distribution in the subset like the whole data.

Since ELGH data lacked variant effect information, we used the Ensembl browser's Variant Effect Predictor (VEP) (McLaren, et al., 2016) tool to fetch VEP annotation for ELGH variants, then merged it with the ELGH data. Our ELGH data got reduced to 3028 records after including VEP annotation and discarding variants with missing VEP annotations.

We further aligned the format of ELGH data with our dataset's schema by dropping the unwanted columns and mapping the ELGH columns to the corresponding gnomAD columns. We considered ELGH's allele frequency of controls as the representative of the healthy population and mapped it to the main allele frequency of the dataset. ELGH's allele count was mapped to both the main allele count and Allele Count South Asian of the dataset. We mapped Homozygote Count from ELGH data with Homozygote Count and Homozygote Count South Asian. We calculated Allele Number and Allele Number South Asian for ELGH data by dividing allele count with allele frequency. We set all allele frequencies for all other ethnicities as zero. Since this data represented the healthy population, we used clinical significance as Benign for all ELGH variants.

Finally, we merged both the Pathogenic and Benign labelled subsets to obtain the final dataset of 5040 variants. Next, we obtained CADD annotations to enrich our dataset further and sourced the CADD database from the Ensembl browser containing 261 million variants. However, owing to the CADD framework's current scope predominantly focusing on SNVs and small indels, the CADD database returned the score only for 102 of the 5040 variants of our dataset, making the CADD annotation unusable for our project.

To minimize the class imbalance in our dataset, we merged Likely pathogenic and Pathogenic/Likely pathogenic classes into the Likely pathogenic class. Figure-1 shows the distribution of clinical significance in our dataset. We were conscious of the remaining imbalance, and we took steps to manage this in preprocessing and modelling steps. We decided against further reducing the dataset to minimize the risk of overfitting.

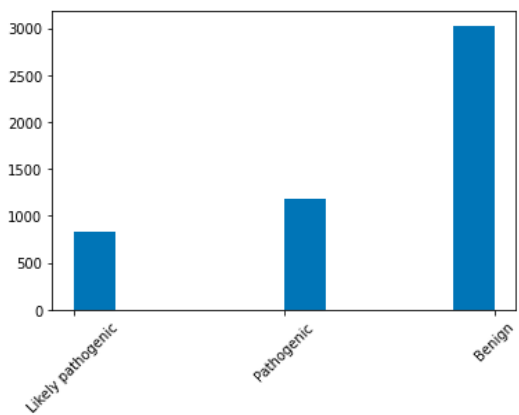

Figure-1: Distribution of clinical significance in the dataset.

To plot the frequency distribution of VEP annotations in our dataset, as shown in Figure-2, we used the binary notion of Benign and Not Benign, where the latter represented both Pathogenic and Likely pathogenic classes. Figure-2 indicates that most Benign variants fall under intron variants, and a tiny number of Benign belong to the other categories. Whereas not many not-Benign variants are intron variants, mainly spread over stop-gained, frameshift, splice donor or acceptor and splice region categories. Probabilistically speaking, this distribution will perhaps cause discriminating Benign variants to be relatively more straightforward for most popular classification algorithms. Intron variants exist in the intron regions of the genes that do not code for the proteins; hence, most intron variants are expected to be benign.

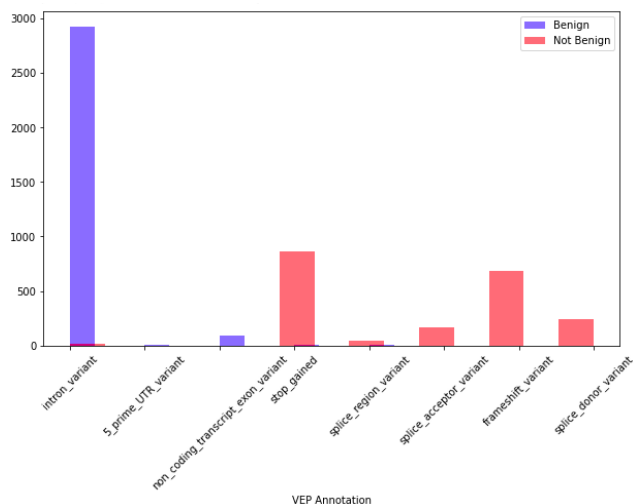

Figure-2: Distribution of variant effects in the dataset.

The allele frequency plot in Figure-3 shows high allele frequencies for Benign labelled variants and very low allele frequencies for Pathogenic/Likely pathogenic variants. We used the binary notion of Benign and Not Benign, where the latter represented both Pathogenic and Likely pathogenic classes. Since Pathogenic LOF variants are expected to be rare and exist in very few individuals, and Benign variants are likely to exist in the majority populations, the distribution of allele frequencies in the data aligns with this norm and indicates good data quality.

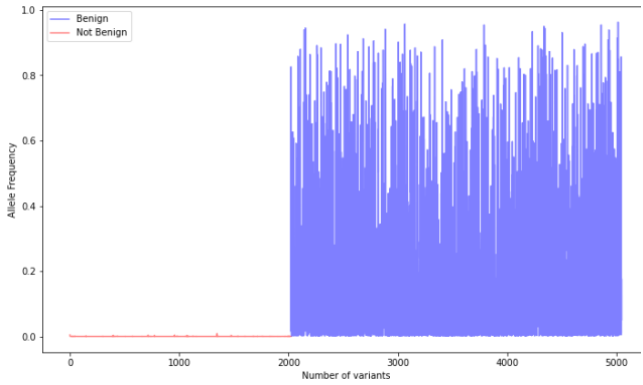

Figure-3: Allele frequency distribution in the dataset.

We have used Matplotlib (Hunter, 2007) to generate the plots.

*B. Preprocessing*

We preprocessed our dataset for preparing training and testing sets using Pandas (McKinney, 2010) (The pandas development team, 2020) and NumPy (Harris, et al., 2020). The initial dataset with 54 attributes got reduced to 41 predictors and one label after we removed the following 12 attributes from the training data considering suitability for training, missing values, and redundancy: Filters - exomes, Filters - genomes, ClinVar Variation ID, Flags, Position, rsIDs, Source, Transcript, HGVS Consequence, Protein Consequence, Transcript Consequence, Allele Number.

We obtained numeric representation of categorical values in Chromosome, VEP Annotation and Clinical Significance using Panda's Factorize method. We got Chromosome transformed to the discrete inclusive range of 0-22, VEP Annotation in the discrete inclusive range of 0-5 and Clinical Significance in the discrete inclusive range of 0-2.

Alternate and Reference alleles had 68 and 178 unique non-numeric values. We applied hash encoding (Weinberger, et al., 2009) to transform these columns to numeric values using the default 8-bit representation of the HashingEncoder class of the Python package category_encoders. Hash encoding resulted in a representation spanning eight columns that increased the number of predictors in the training data to 47. Table-1 shows the list of all predictors in the training data.

To better understand the generalizability of our models, we set aside about six per cent (300 records) of the data before further preprocessing and training to test the models after training and validation. We sampled the test data keeping the chromosome distribution like the whole training data.

To address the class imbalance in training data, we created another version of the training data by applying Synthetic Minority Over-sampling Technique (SMOTE) (Chawla, et al., 2002) using the imblearn Python library (Lemaître, et al., 2017). SMOTE augments the data by synthesizing samples for the minority classes and results in a perfectly balanced dataset. After applying SMOTE, another version of our training data contained 8526 records in total or 2842 records per class.

| Column No. | Description | Column No. | Description |
|---|---|---|---|
| 1 | col_0 (Part1 - Hash Encoded Reference and Alternate Allele) | 25 | Homozygote Count Ashkenazi Jewish |
| 2 | col_1 (Part2 - Hash Encoded Reference and Alternate Allele) | 26 | Hemizygote Count Ashkenazi Jewish |
| 3 | col_2 (Part3 - Hash Encoded Reference and Alternate Allele) | 27 | Allele Count East Asian |
| 4 | col_3 (Part4 - Hash Encoded Reference and Alternate Allele) | 28 | Allele Number East Asian |
| 5 | col_4 (Part5 - Hash Encoded Reference and Alternate Allele) | 29 | Homozygote Count East Asian |
| 6 | col_5 (Part6 - Hash Encoded Reference and Alternate Allele) | 30 | Hemizygote Count East Asian |
| 7 | col_6 (Part7 - Hash Encoded Reference and Alternate Allele) | 31 | Allele Count European (Finnish) |
| 8 | col_7 (Part8 - Hash Encoded Reference and Alternate Allele) | 32 | Allele Number European (Finnish) |
| 9 | Chromosome | 33 | Homozygote Count European (Finnish) |
| 10 | VEP Annotation | 34 | Hemizygote Count European (Finnish) |
| 11 | Allele Count | 35 | Allele Count European (non-Finnish) |
| 12 | Allele Frequency | 36 | Allele Number European (non-Finnish) |
| 13 | Homozygote Count | 37 | Homozygote Count European (non-Finnish) |
| 14 | Hemizygote Count | 38 | Hemizygote Count European (non-Finnish) |
| 15 | Allele Count African/African-American | 39 | Allele Count Other |
| 16 | Allele Number African/African-American | 40 | Allele Number Other |
| 17 | Homozygote Count African/African-American | 41 | Homozygote Count Other |
| 18 | Hemizygote Count African/African-American | 42 | Hemizygote Count Other |
| 19 | Allele Count Latino/Admixed American | 43 | Allele Count South Asian |
| 20 | Allele Number Latino/Admixed American | 44 | Allele Number South Asian |
| 21 | Homozygote Count Latino/Admixed American | 45 | Homozygote Count South Asian |
| 22 | Hemizygote Count Latino/Admixed American | 46 | Hemizygote Count South Asian |
| 23 | Allele Count Ashkenazi Jewish | 47 | oe_lof_upper |
| 24 | Allele Number Ashkenazi Jewish | | |

Table-1: Predictors in the training data

However, the SMOTE method generated decimal values for Chromosome and resulted in 2923 unique values. To fix this, we rounded off the Chromosome values to the nearest integers, which restored the 23 unique values for Chromosome.

We created another version of the training data by performing Principal Component Analysis (PCA). We used the PCA class of Scikit-learn (Pedregosa, et al., 2011) instantiated with 0.95 to capture 95% variance of the data. PCA resulted in 17 principal components. Figure-4 shows the heatmap of each principal component capturing the degree of variance from each feature of the training data. We used Seaborn (Waskom, 2021) to generate the heatmap.

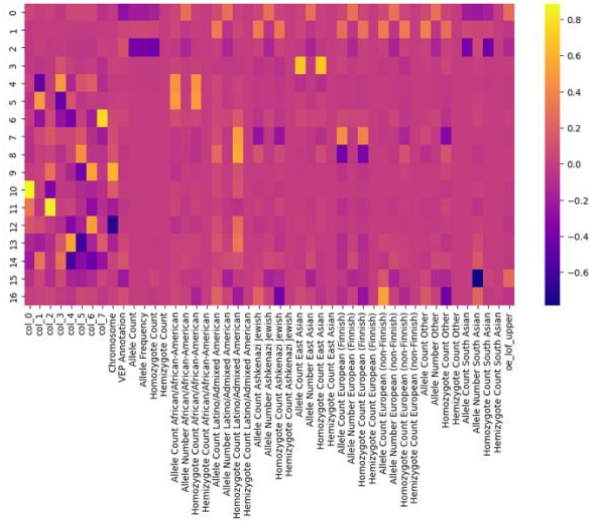

Figure-4: Heatmap showing the extent of variance captured by principal components for each feature in the training data.

To standardize all the training data features across all versions, we applied z-score normalization on each feature by subtracting the feature's mean across all samples and dividing by the feature's standard deviation.

### C. Training, Validation and Testing

We split the training data into training and testing sets, with the training set containing 70% of the data. We trained seven classification algorithms, including Logistic Regression, K-Nearest Neighbors (KNN), Support Vector Machine (SVM), Naïve Bayes and Random Forest Classifier (RFC) from Scikit-learn (Lemaître, et al., 2017), XGBoost (Chen & Guestrin, 2016) and TabNet (Arık & Pfister, 2020).

For Logistic Regression, we used L2-regularization, LBFGS optimizer and max iterations = 20000 to ensure the convergence.

For tuning the hyperparameters of the SVM model, we performed an automated grid search using Scikit-learn's GridSearchCV class. We used the best parameter results returned by the grid search for the training.

We manually tuned the value of K for the KNN algorithm and settled with nine neighbours after multiple runs. Additionally, we used leaf size 30 and Minkowski distance as the metric.

We did not specify any prior for the Gaussian Naïve Bayes algorithm and let the Scikit-learn determine the priors based on the data. We used variance smoothing as 0.00001 after manual tuning based on multiple runs.

Random Forest classifier (RFC) is a machine learning ensemble algorithm based on the bagging mechanism that aggregates the results of multiple decision trees. For the RFC, we found the number of estimators/trees in the forest as 50 to be the most effective for our data after multiple manual iterations. Additionally, we used the Gini Impurity criterion for measuring the split quality, max number of features to consider for the best fit as the square root of the feature count, and whole data bootstrapping for building the trees.

XGBoost, a relatively new machine learning ensemble algorithm based on boosting mechanism, uses gradient boosted trees, runs with more efficient time and space complexity and has been more successful than many other standalone and ensemble algorithms (Chen & Guestrin, 2016). We ran an automated grid search using GridSearchCV class of Scikit-learn to find the best-tuned hyperparameters for the XGBoost algorithm. Additionally, we used the objective function as "multi:softmax" for multi-class classification and scoring criterion as "roc_auc_ovr_weighted" to evaluate the candidate models using the area under the receiver operating curve with weighted average and one vs rest multi-class evaluation.

We also trained the TabNet algorithm implemented in the class TabNetClassifier of the package pytorch_tabnet, an attention mechanism based deep learning algorithm for tabular data (Arık & Pfister, 2020).

We trained all seven algorithms on three different versions of our training sets and calculated testing accuracies using testing sets. Due to the multi-class data, we considered testing accuracy an insufficient measure of the model performance; hence we also calculated metrics such as Precision, Recall, Confusion Matrix and F1-score for better visibility of the class-level performance. We further selected the two best performing models according to the F1-score, trained them on the entire training data, and tested their performance on the independent test set saved before training to gain further insight on the generalizability of the models.

As an additional experiment, we created several ensembles from different combinations of the seven algorithms we trained and evaluated their performance against the best performing individual classifiers. We used three different methods based on the weighted majority vote, highest average class wise probability, and stacking for creating ensembles using Scikit-learn's sklearn.ensemble module.

### IV. RESULTS

#### A. Testing Set

Upon training with the first/original version of the training data, all the models except Naïve Bayes scored testing accuracy beyond 0.8, whereas XGBoost (XGB), RFC, KNN and SVM scored testing F1-score above 0.8. XGB and RFC proved to be the two best models, respectively, according to the F1 score. Figure-5 shows the comparison plot of F1-scores of all models on the testing set of the original version of the training data.

Further evaluation using the confusion matrix in Figure-6 and Figure-7 suggests that both the models significantly misclassify the Likely Pathogenic class and score 100% accuracy on the Benign class. Table-2 summarizes all the metrics viz. testing accuracy, F1-score, precision and recall for all the models on the testing set of the original version of the training data.

We used different combinations of ensembles such as LR-SVM-KNN-NB, LR-SVM-KNN, LR-KNN-SVM-RFC-XGB, SVM-RFC-XGB and RFC-XGB. Our ensembles used three different methods viz. majority weighted voting, highest average probability per class and stacking (with XGB being the final estimator and others in the combination being the initial estimators). We observed the best results with the stacking ensemble using RFC as the initial and XGB as the final estimator. However, even the best performing ensemble could not exceed the performance of the individual XGB model.

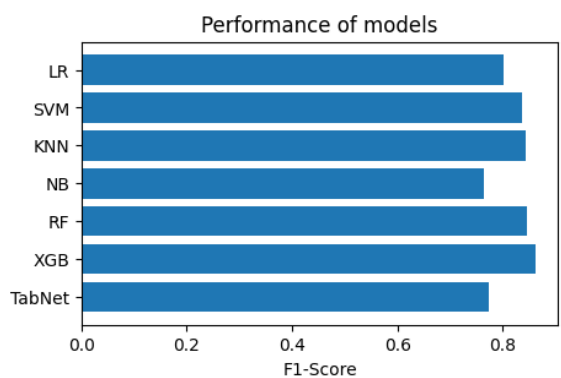

Figure-5: F1-score comparison on the testing set of the original version of the training data

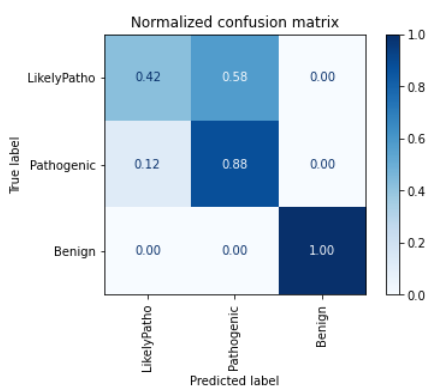

Figure-6: Confusion Matrix of XGBoost model on the testing set of the original version of the training data

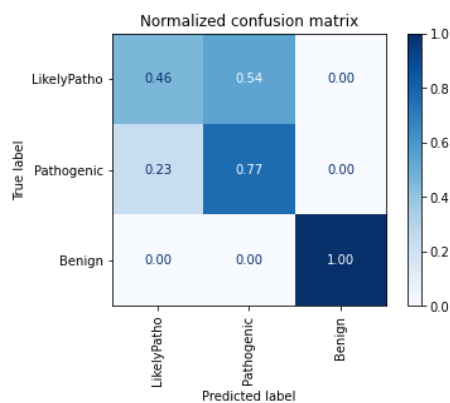

Figure-7: Confusion Matrix of RFC model on the testing set of the original version of the training data

The testing set results of the SMOTE version of the training data show that both RFC and XGB models improve significantly on the Likely Pathogenic class, as shown in Figure-8 and 9.

| Dataset Version | Model | Testing | | | | |
| --- | --- | --- | --- | --- | --- | --- |
| | | Testing Accuracy | F1-Score | Class | Precision | Recall |
| Original | LR | 0.82 | 0.8 | Likely Path. | 0.48 | 0.18 |
| | | | | Pathogenic | 0.6 | 0.86 |
| | | | | Benign | 1 | 1 |
| | SVM | 0.84 | 0.84 | Likely Path. | 0.55 | 0.43 |
| | | | | Pathogenic | 0.65 | 0.75 |
| | | | | Benign | 1 | 1 |
| | KNN | 0.85 | 0.84 | Likely Path. | 0.59 | 0.43 |
| | | | | Pathogenic | 0.66 | 0.79 |
| | | | | Benign | 1 | 1 |
| | NB | 0.79 | 0.76 | Likely Path. | 0.44 | 0.87 |
| | | | | Pathogenic | 0.71 | 0.22 |
| | | | | Benign | 1 | 1 |
| | RFC | 0.85 | 0.85 | Likely Path. | 0.58 | 0.46 |
| | | | | Pathogenic | 0.67 | 0.77 |
| | | | | Benign | 1 | 1 |
| | XGB | 0.87 | 0.86 | Likely Path. | 0.71 | 0.42 |
| | | | | Pathogenic | 0.68 | 0.88 |
| | | | | Benign | 1 | 1 |
| | TabNet | 0.83 | 0.77 | Likely Path. | 0.78 | 0.03 |
| | | | | Pathogenic | 0.51 | 0.99 |
| | | | | Benign | 1 | 1 |

Table-2: Testing set results for all the models on the original version of the training data

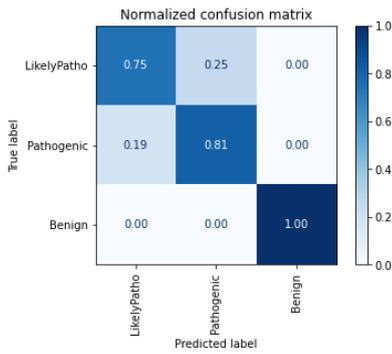

Figure-8: Confusion Matrix of XGBoost model on the testing set of the SMOTE version of the training data

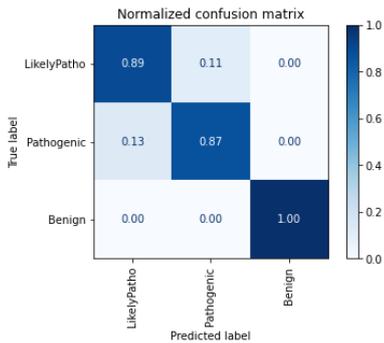

Figure-9: Confusion Matrix of RFC model on the testing set of the SMOTE version of the training data

Table-3 summarizes all the metrics viz. testing accuracy, F1-score, precision and recall for all the models on the testing set of the SMOTE version of the training data. Results show that XGB, RFC and SVM improve substantially on the Likely Pathogenic class, which got heavily misclassified on the original version of the training data.

| Dataset Version | Model | Testing | | | | |
| --- | --- | --- | --- | --- | --- | --- |
| | | Testing Accuracy | F1-Score | Class | Precision | Recall |
| SMOTE | LR | 0.71 | 0.71 | Likely Path. | 0.55 | 0.65 |
| | | | | Pathogenic | 0.6 | 0.5 |
| | | | | Benign | 1 | 1 |
| | SVM | 0.83 | 0.83 | Likely Path. | 0.73 | 0.76 |
| | | | | Pathogenic | 0.76 | 0.74 |
| | | | | Benign | 1 | 1 |
| | KNN | 0.77 | 0.77 | Likely Path. | 0.64 | 0.66 |
| | | | | Pathogenic | 0.68 | 0.65 |
| | | | | Benign | 1 | 1 |
| | NB | 0.66 | 0.61 | Likely Path. | 0.49 | 0.88 |
| | | | | Pathogenic | 0.58 | 0.15 |
| | | | | Benign | 1 | 1 |
| | RFC | 0.92 | 0.92 | Likely Path. | 0.87 | 0.89 |
| | | | | Pathogenic | 0.9 | 0.87 |
| | | | | Benign | 1 | 1 |
| | XGB | 0.85 | 0.85 | Likely Path. | 0.79 | 0.75 |
| | | | | Pathogenic | 0.78 | 0.81 |
| | | | | Benign | 1 | 1 |
| | TabNet | 0.76 | 0.76 | Likely Path. | 0.66 | 0.58 |
| | | | | Pathogenic | 0.65 | 0.72 |
| | | | | Benign | 1 | 1 |

Table-3: Testing set results for all the models on the SMOTE version of the training data

Likely pathogenic class also gets notably misclassified on the PCA version of the data, as shown by the testing set results in Table-4 and Figure-11. The SVM model scores slightly better F1-score on the PCA data than the RFC and XGB models, as Figure-10 suggests.

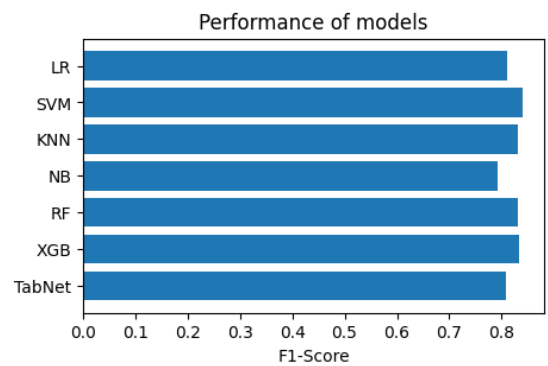

Figure-10: F1-score comparison of the models on the testing set of the PCA data.

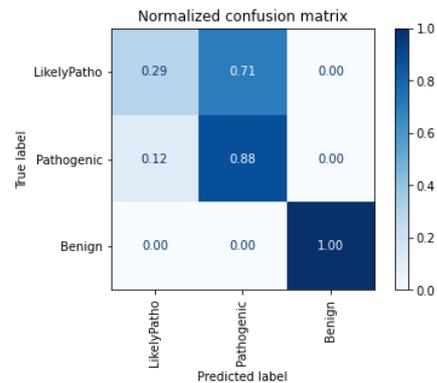

Figure-11: Confusion matrix of the XGB model on the testing set of the PCA data

| Dataset Version | Model | Testing | | | | |
| --- | --- | --- | --- | --- | --- | --- |
| | | Testing Accuracy | F1-Score | Class | Precision | Recall |
| PCA | LR | 0.84 | 0.81 | Likely Path. | 0.67 | 0.18 |
| | | | | Pathogenic | 0.59 | 0.92 |
| | | | | Benign | 1 | 1 |
| | SVM | 0.84 | 0.84 | Likely Path. | 0.53 | 0.42 |
| | | | | Pathogenic | 0.63 | 0.76 |
| | | | | Benign | 1 | 1 |

| Dataset Version | Model | Testing | | | | |
|---|---|---|---|---|---|---|
| | | Testing Accuracy | F1-Score | Class | Precision | Recall |
| | KNN | 0.84 | 0.83 | Likely Path. | 0.58 | 0.34 |
| | | | | Pathogenic | 0.61 | 0.81 |
| | | | | Benign | 1 | 1 |
| | NB | 0.82 | 0.79 | Likely Path. | 0.51 | 0.14 |
| | | | | Pathogenic | 0.57 | 0.89 |
| | | | | Benign | 1 | 1 |
| | RFC | 0.84 | 0.83 | Likely Path. | 0.55 | 0.4 |
| | | | | Pathogenic | 0.61 | 0.74 |
| | | | | Benign | 1 | 1 |
| | XGB | 0.85 | 0.83 | Likely Path. | 0.66 | 0.29 |
| | | | | Pathogenic | 0.61 | 0.88 |
| | | | | Benign | 1 | 1 |
| | TabNet | 0.83 | 0.8 | Likely Path. | 0.6 | 0.19 |
| | | | | Pathogenic | 0.58 | 0.9 |
| | | | | Benign | 1 | 1 |

Table-4: Testing set results for all the models on the PCA version of the training data

*B. Independent Test Set (ITS)*

Out of the 21 models trained during the training and validation phase, we selected six models, XGB and RFC, from each of the three versions of the data. We trained these selected models using the corresponding versions' entire training data (training + testing sets). Results show that the models trained on the SMOTE version perform poorly on the ITS, whereas those trained on the original version show the best performance. However, the performance on the Likely pathogenic class remains relatively poor. Table-5 summarizes the results on the ITS.

| Dataset Version | Model | Testing with the independent test set | | | | |
|---|---|---|---|---|---|---|
| | | Testing Accuracy | F1-Score | Class | Precision | Recall |
| Original | RFC | 0.86 | 0.86 | Likely Path. | 0.53 | 0.44 |
| | | | | Pathogenic | 0.69 | 0.76 |
| | | | | Benign | 1 | 1 |
| | XGB | 0.89 | 0.88 | Likely Path. | 0.68 | 0.44 |
| | | | | Pathogenic | 0.72 | 0.87 |
| | | | | Benign | 1 | 1 |
| SMOTE | RFC | 0.2 | 0.12 | Likely Path. | 0.05 | 0.19 |
| | | | | Pathogenic | 0.35 | 0.75 |
| | | | | Benign | 0 | 0 |
| | XGB | 0.17 | 0.13 | Likely Path. | 0.1 | 0.58 |
| | | | | Pathogenic | 0.59 | 0.37 |
| | | | | Benign | 0 | 0 |

| Dataset Version | Model | Testing with the independent test set | | | | |
|---|---|---|---|---|---|---|
| | | Testing Accuracy | F1-Score | Class | Precision | Recall |
| PCA | RFC | 0.83 | 0.83 | Likely Path. | 0.47 | 0.37 |
| | | | | Pathogenic | 0.62 | 0.75 |
| | | | | Benign | 1 | 1 |
| | XGB | 0.85 | 0.81 | Likely Path. | 0.4 | 0.09 |
| | | | | Pathogenic | 0.63 | 0.9 |
| | | | | Benign | 0.99 | 1 |

Table-5: Independent test set results of the XGB and RFC models trained on different versions of the data

## V. DISCUSSION

We curated a novel dataset using some high-quality databases available for human genetic variation. We trained several classification algorithms, and our eventually selected model demonstrated adequate performance overall, with an F1-score of 0.88 on the independent test data. However, the performance on the Likely pathogenic class remains subpar. The classification models find it hard to discriminate Likely pathogenic from Pathogenic samples. As the results show, all the misclassified Likely pathogenic samples get classified as Pathogenic. Combining the Pathogenic and Likely pathogenic classes seems likely to improve the performance further, but that may dilute the model's utility for the use cases requiring prioritization of potentially pathogenic LOF variants for clinical reasons such as drug targeting.

Nevertheless, the current model shows high recall/sensitivity for the Pathogenic class and moderate precision for Likely pathogenic samples. It would be interesting to test the model in a more realistic environment and compare it with the currently used methods in the real world. Since the LOF variants have not been sufficiently studied compared to SNPs, not all the valuable annotations such as CADD currently exist for LOFs. Our approach demonstrated a fair relevance of machine learning methods in classifying LOFs. As more LOF data and more effective annotations become available for LOF variants, we foresee machine learning methods proving to be more effective.

## VI. FUTURE WORK

Several out-of-the-box machine learning classification algorithms available today are reasonably mature and work well for most use cases. Though exploring new machine learning methods specifically for LOFs and genetic variation seems a possible future direction, focusing more on further data enrichment will likely yield better results. Including some gene-level annotations such as conservation score and exploring the possibility of linking cross-species genes data with the LOF variants might further improve the data quality. Bringing in some protein-based predictors in the dataset also appears a fine prospect. As applicable to most projects dealing with genomic data, further upgrading the dataset as more data and findings become available is definite future work.